# BUILDING TOFFOLI NETWORK FOR REVERSIBLE LOGIC SYNTHESIS BASED ON SWAPPING BIT STRINGS

Hafiz Md. Hasan Babu, Md. Saiful Islam, Md. Rafiqul Islam, Lafifa Jamal, Abu Ahmed Ferdaus, Muhammad Rezaul Karim And Abdullah Al Mahmud

Department of Computer Science & Engineering, University of Dhaka
Dhaka-1000, Bangladesh
Email: {hafizbabu@hotmail.com}, {sohel_csdu, rafik3203, r_karimcs, aamrubel}@yahoo.com

**Abstract**

In this paper, we have implemented and designed a sorting network for reversible logic circuits synthesis in terms of *n\*n* Toffoli gates. The algorithm presented in this paper constructs a Toffoli Network based on swapping bit strings. Reduction rules are then applied by simple template matching and removing useless gates from the network. Random selection of bit strings and reduction of control inputs are used to minimize both the number of gates and gate width. The method produces near optimal results for up to 3-input 3-output circuits.

## I. Introduction

Landaur's principle[1] proved that logic computations that are not reversible necessarily dissipate heat irrespective of their implementation technologies. Bennet[2] showed that zero energy dissipation would possible only if the network consists of reversible gates. Thus reversibility will become an essential property in future circuit design.

Synthesis of reversible logic circuits differs significantly from the synthesis of combinational (classical) logic circuits. Because in a reversible circuit the number of inputs must be equal to the number of outputs, every output can be used only once (i.e., no fan-out is permitted), and must be acyclic.

Although there exist many reversible gates in the literature good synthesis methods have not yet emerged. Miller *et al.*[8] used spectral techniques to find near optimal circuits. Mischenko and Perkowski[6] suggested a regular structure of reversible wave cascades and have shown that such a structure requires no more than product terms in an ESOP realization of the function. Miller *et al.*[9] have developed a transformation-based algorithm for reversible logic synthesis. A regular symmetric structure has been proposed by Perkowski *et al.*[7] to realize symmetric functions. In fact one would expect that a better method can be found.

In this paper, we have given algorithms that synthesize the circuit in one direction. The algorithms build a network consisting of a sequence of Toffoli gates read from left to right. The applications of the algorithm in both input and output translations have been described. Template matching and identifying useless gates are used to reduce both the number of gates and circuit width.

## II. Background

In this section we provide some definitions and background on reversible logic.

***Definition 2.1*** An n-input n-output totally specified Boolean function $f(X)$, $X = \{x_1, x_2, …, x_n\}$ is reversible iff it maps each input assignment to a unique output assignment.

A reversible function can be written as a standard truth table as in Table 2.1 and can also be viewed as a bijective mapping of the set of integers $0,1, …, 2^n-1$. Hence a reversible function can be defined as an ordered set of integers corresponding to the right side of the table, e.g. $\{1,0,3,2,5,7,4,6\}$ for the function in Table 2.1. We can thus interpret the function over the integers as $f(0) = 1$, $f(1) = 0$, $f(2) = 3$, etc.

| c | b | a | c° | b° | a° |
|---|---|---|----|----|----|
| 0 | 0 | 0 | 0  | 0  | 1  |
| 0 | 0 | 1 | 0  | 0  | 0  |
| 0 | 1 | 0 | 0  | 1  | 1  |
| 0 | 1 | 1 | 0  | 1  | 0  |
| 1 | 0 | 0 | 1  | 0  | 1  |
| 1 | 0 | 1 | 1  | 1  | 1  |
| 1 | 1 | 0 | 1  | 0  | 0  |
| 1 | 1 | 1 | 1  | 1  | 0  |

**Table 2.1 3*3 Reversible Logic Function**

***Definition 2.2*** An n-input n-output gate is reversible if it realizes a reversible function.

Many reversible gates have been proposed in the literature. One of the first gates was the CNOT gate (Feynman[3]), which capable of producing the "exclusive or" of two input bits as the second output and the first output is equal to the first input. A generalization of CNOT is a 3-input 3-output Toffoli gate (Toffoli[4]). The Toffoli gate negates the third bit iff the first two bits are 1. Figure 2.1 shows both gates as they are commonly drawn.

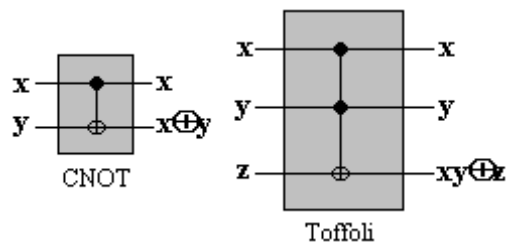

**Fig 2.1 CNOT and Toffoli gates**

In this paper, we use the generalized family of Toffoli gates defined as follows:

***Definition 2.3*** A generalized $n*n$ Toffoli gate changes one bit, called the target, if some of the *k* bits are 1(Figure 2.2). The changing bit (target) may also be in any position.

The gate will be defined as follows TOF($x_{i1}, x_{i2}, …, x_{ik}; x_n$) where $x_n$ is the target and $x_{i1}, x_{i2}, …, x_{ik}$ are the control bits.

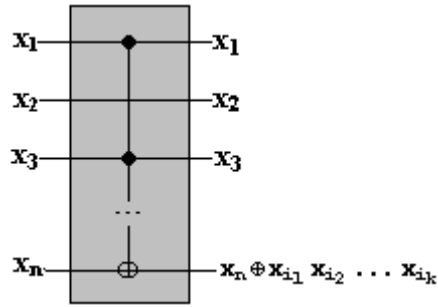

**Fig 2.2 Generalized Toffoli gate**

***Definition 2.4*** Given two bit strings, *P* and *Q*, the Hamming distance between them, denoted $\delta(P, Q)$ is the number of positions for which *P* and *Q* differ.

***Lemma 2.1*** In any reversible specification the upper and lower bound on the Hamming distance between any two bit strings *P* and *Q*, is *n* and 1, where n is the number of input lines. That is

$$1 \leq \delta(P, Q) \leq n$$

*Proof:* Let *P* be $(a_1, a_2, …, a_n)$ and *Q* be $(b_1, b_2, …, b_n)$. Since in a reversible specification no two bit strings are identical, they must differ in at least one position. Let *m* be the index at which $a_m \neq b_m$. That is, $a_m = b_m'$, and $b_m$ is either 0 or 1. Bit strings *P* and *Q* may differ at most every position, i.e., $a_i \neq b_i$, where $1 \leq i \leq n$. Thus we can conclude that $1 \leq \delta(P, Q) \leq n$.

***Definition 2.5*** Given the function $f(X)$, the complexity $C(f)$ is defined as the sum of the individual Hamming distances over the $2^n$ input-output patterns.

For example, the value of $C(f)$ for the function in Table 2.1 is 8.

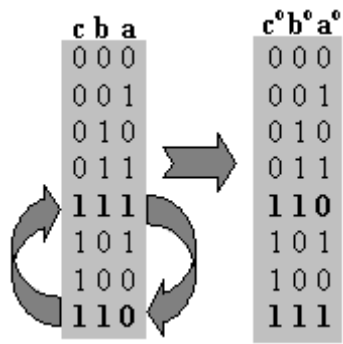

**Fig 2.3 Swapping Bit Strings**

***Lemma 2.2*** Two bit strings *P* and *Q* can be swapped without affecting others iff the Hamming distance between them is 1.

For example bit strings (1,1,1) and (1,1,0) have been swapped using TOF(*b*,*c*;*a*), which is shown in Figure 2.3. Since the Hamming distance between them is one, this swapping can be carried out without affecting others.

### III. Building Network

We have already described that a reversible function can be defined as an ordered set of integers corresponding to the right side of the table, e.g. {*1,0,3,2,5,7,4,6*} for the function in Table 2.1. Therefore, if we can build a network of reversible gates that might sort this set, it will eventually realize the function. For example, the ordered set of integers for the function in table 2.1 will become {*0,1,2,3,4,5,6,7*} and the index of each element will be equal to itself. That is, if we define the set as $\{p_i\}$, then $2^n-1 \geq i \geq 0$ and $p_i = i$ where *n* is the number of input lines. In this paper we have presented two algorithms regarding this. The network is build as a sequence of Toffoli gates from the output side to the input side. Of course, applying the sequence in reverse transforms the input side to the output side.

### Output Translation

To begin, we present algorithms that identify Toffoli gates only on the output side of the specification.

### Algorithm_1

**Step 1:** Take the first bit string $(a_1,a_2,...,a_n)$ in the set and bring it to its intended place. If its index is equal to itself, then we are done and we can start over from step 4. If it is not, by induction, its place is occupied by another string, say $(b_1,b_2,..., b_n)$.

**Step 2:** Compute the Hamming distance between $(a_1,a_2,...,a_n)$ and $(b_1,b_2,...,b_n)$. If distance is one, simply swap them and goto step 4. If distance is more than one, goto step 3.

**Step 3:** Find all the bit strings $(c_1,c_2, ... ,c_n)$ such that the Hamming distance between $(b_1,b_2,...,b_n)$ and $(c_1,c_2, ... ,c_n)$ is one. Swap $(b_1,b_2,...,b_n)$ with one of the $(c_1,c_2,...,c_n)$ so that the Hamming distance between $(a_1,a_2,...,a_n)$ and $(c_1,c_2,...,c_n)$ is minimum. If more than one such $(c_1,c_2,...,c_n)$ is found take the one whose integer representation is low and is not in its intended place. Now $(c_1,c_2,...,c_n)$ will become the new $(b_1,b_2,...,b_n)$ and goto step 2.

**Step 4:** If all the bit stings in the set in its intended place, algorithm finishes. Otherwise goto step 1.

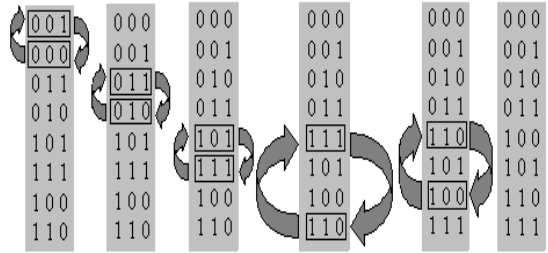

**Fig 3.1 Building Network using Algorithm_1**

Figure 3.1 illustrates the application of the algorithm_1. The sequence of gates that form the network is: TOF($b'$,$c'$;$a$)TOF($b$,$c'$;$a$)TOF($a$,$c$;$b$)TOF($b$,$c$;$a$)TOF($a'$,$c$;$b$).

Algorithm_1 is straightforward. It is greedy in the sense it hopes that a bit string can be swapped by the one that is in its intended place. Because of lemma 2.1 it is always possible to find two bit strings for step 2 and 3 that can be swapped. Therefore, it will always terminate successfully with a circuit for the given specification. However, it is always possible to find a circuit to realize a function that requires at most $n \times 2^n$ gates. The best case occurs when a bit string can always be swapped by the bit string placed in its intended position.

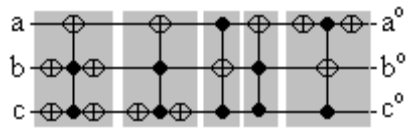

**Fig 3.2 Circuit for the Network Build by Algorithm_1**

## Algorithm_2

Algorithm_2 differs from algorithm_1 in step 1 only. Algorithm_2 takes the bit string in the set whose integer representation is low and brings it to its intended place. Steps 2, 3, 4 of algorithm_1 are the same for algorithm_2.

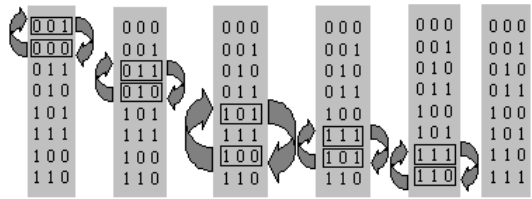

**Fig 3.3 Building Network using Algorithm_2**

Figure 3.3 illustrates the application of the algorithm_2 and the sequence of gates that form the network is: TOF($b'$,$c'$;$a$) TOF($b$,$c'$;$a$) TOF($b'$,$c$;$a$) TOF($a$,$c$;$b$) TOF($b$,$c$;$a$). Like algorithm_1, algorithm_2 is also greedy and hopes that a bit string can always be swapped by the one that is in its intended place.

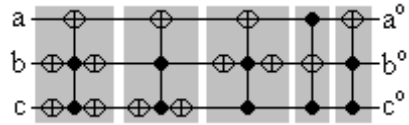

Fig 3.4 Circuit for the Network Build by Algorithm_2

**Input Translation**

For input translation we have to find an inverse of the specification. For example, the reverse specification of the function in table 2.1 is {1,0,3,2,6,4,7,5} as shown in Figure 3.5. Then we can apply algorithm_1 and algorithm_2 to realize the function.

Applying algorithm_1 to realize the reverse specification we get the circuit: TOF($b',c';a$) TOF($b,c';a$) TOF($b,c;a$) TOF($a,c;b$) TOF($b',c;a$). Applying algorithm_2 we get: TOF($b',c';a$) TOF($b,c';a$) TOF($a',c;b$) TOF($b,c;a$) TOF($a,c;b$) T($b,c;a$).

**Random Selection and Control Input Reduction**

To sort the elements in a specification algorithm_1 and algorithm_2 take one element at a time and brings it to its intended place. Heuristic approaches can be applied to select the elements so that the total number of swapping is reduced. This will minimize the total number of gates in the network also.

| Function $f$ | | Function $f^{-1}$ | |
|---|---|---|---|
| c b a | c°b°a° | c°b°a° | c b a |
| 0 0 0 | 0 0 1 | 0 0 0 | 0 0 1 |
| 0 0 1 | 0 0 0 | 0 0 1 | 0 0 0 |
| 0 1 0 | 0 1 1 | 0 1 0 | 0 1 1 |
| 0 1 1 | 0 1 0 | 0 1 1 | 0 1 0 |
| 1 0 0 | 1 0 1 | 1 0 0 | 1 1 0 |
| 1 0 1 | 1 1 1 | 1 0 1 | 1 0 0 |
| 1 1 0 | 1 0 0 | 1 1 0 | 1 1 1 |
| 1 1 1 | 1 1 0 | 1 1 1 | 1 0 1 |

Fig 3.5 Reverse Specification

Algorithm_1 and algorithm_2 also assigns the maximum number of control lines to each Toffoli gate. For larger problems with up to 8 or 9 inputs this may not be a practical one. Selective use of control inputs can be used to swap elements. This can be carried out safely as long as it will not affect bit strings that are already in its intended place. We should choose a subset of the control inputs that will minimize the C($f$) of the resulting specification. For example, we can select control inputs that will drive a Toffoli gate to bring more than one element at a time to their intended place.

**Reduction Rules**

The circuits produced by the algorithm as described thus far frequently have gate sequences that can be reduced. For example, the sequence TOF(*b*;*a*) TOF(;*b*) TOF(;*a*) can be replaced by the sequence TOF(;*b*) TOF(*b*;*a*). Here we have implemented template driven reduction method introduced Miller *et al.*[9]. A template consists of a sequence of gates to be matched and the sequences of gates to be substituted when a match is found. The lines in the template are generic and must be associated to real lines in the circuit. The template matching procedure looks for the target gates, including the initial match to the widest gate, across the entire circuit. If all target gates are found, it attempts to move the gates so that they are adjacent either matching the template in the forward or reverse direction. If this can be done, the matched gates are replaced with the new gates specified by the template.

When moving the target gates, the matching procedure takes account of Property 4.1 which follows directly from the definition of *n\*n* Toffoli gates. If two gates cannot be interchanged because they do not satisfy this property and that prohibits proper adjacent ordering of the target gates for a match, the template being considered is not applicable.

**PROPERTY 4.1** Two gates TOF($x_1,x_2,…,x_{k-1}$ ; $x_k$) and TOF($y_1,y_2,…,y_{l-1}$; $y_l$) adjacent in a circuit can be interchanged iff $x_k$ not in { $y_1,y_2,…,y_{l-1}$ } and $y_l$ not in { $x_1,x_2,…,x_{k-1}$ }.

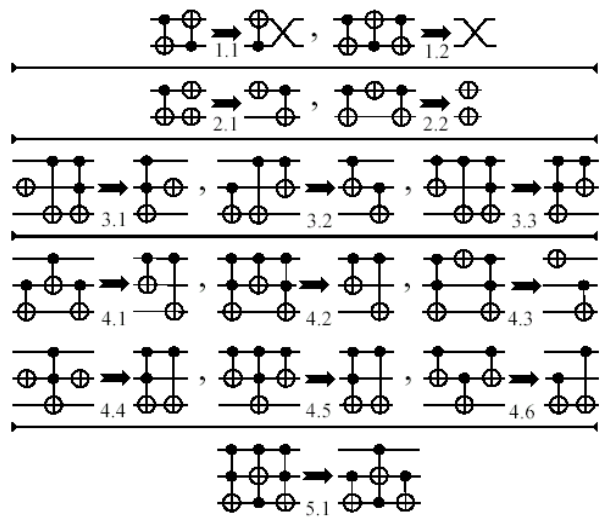

Fig 4.1 Templates with 2 or 3 inputs

We have also removed useless gates that come in pairs and have no effect in the circuit. For example, TOF(a,b;c) is an example of a useless gate in the gate sequence TOF(a,b;c)TOF(a;c)TOF(a,b;c) and the gate sequence can be replaced by TOF(a;c) without any modification in the circuit.

When identifying useless gate in a circuit, we have taken into account of Property 4.2 which follows directly from the definition of *n\*n* Toffoli gates.

**PROPERTY 4.2** A gate TOF($x_1,x_2,…,x_{k-1}; x_k$) can be removed from the sequence TOF($x_1,x_2,…,x_{k-1}; x_k$) TOF($a_1,a_2,…,a_{l-1}; a_l$)TOF($b_1,b_2,…,b_{m-1};b_m$)…TOF($c_1,c_2,… ,c_{n-1}; c_n$) TOF($x_1,x_2,…,x_{k-1}; x_k$) iff $x_k$ not in { $a_1,a_2,…, a_{l-1}, b_1,b_2,…,b_{m-1},…,c_1,c_2,… ,c_{n-1}$} and $a_l,b_m,…,c_n$ not in { $x_1,x_2,…,x_{k-1}$}.

## IV. Experimental Results

For each example, the specification is given as an ordered set of integers, which define the truth table specification of the reversible logic function to be realized. The circuit is given as an ordered sequence of Toffoli gates. Read from left to right they transform the left side to the right side.

**Example 5.1 Verification** of realizing a **Fredkin gate.** This example is collected from[8]. The circuit given by our method produces the same result.

**Specification:** {0,1,2,3,4,6,5,7}

**Circuit for Algorithm_1:**

TOF(*a,c;b*)TOF(*b,c;a*)TOF(*a,c;b*) $\Rightarrow$ TOF(*a;b*)TOF(*b,c;a*)TOF(*a;b*)

**Circuit for Algorithm_2:**

TOF(*b,c;a*)TOF(*a,c;b*)TOF(*b,c;a*) $\Rightarrow$TOF(*b;a*)TOF(*a,c;b*)TOF(*b;a*)

**Example 5.2** This is a second example of the interchange of two positions in the specification. The circuit given by our method is identical to the solution provided by Perkowski [10].

**Specification:** {0,1,2,4,3,5,6,7}

**Circuit for Algorithm_1:** TOF(*a,b;c*)TOF(*a,c;b*) TOF(*b′,c;a*) TOF(*a,c;b*)(*a,b;c*)

**Circuit for Algorithm_2:** TOF(*a′,b′;c*) TOF(*b′,c′;a*) TOF(*a,c′;b*) TOF(*b′,c′;a*) TOF(*a′,b′;c*)

**Example 5.3** This example is taken from[9]. The circuit given by our method is identical to the solution provided by the Bidirectional Algorithm in[9].

**Specification:** {7,0,1,2,3,4,5,6}

**Circuit for Algorithm_2:** TOF(*a,b;c*)TOF(*a,c′;b*)TOF(*b′,c′*;a) TOF(*b,c′;a*) TOF(*a,c;b*) TOF(*b′,c;a*) TOF(*b,c;a*) $\Rightarrow$ TOF(*a,b;c*)TOF(*a;b*)TOF(;*a*)

Though the initial circuit produced by our algorithm seems to be larger one, it can be simplified easily by simple template matching and identifying useless gates. Thus the circuit will be optimal. The main advantage of our algorithm is that it does not require exhaustive analysis like spectral used in[8].

## V. Conclusion

A very simple but powerful algorithm to realize totally specified reversible specification has been presented. The algorithm will always terminate with a network of Toffoli gates that can translate both input and output side to their corresponding output and input side. Since the synthesis of reversible circuits can be done in either side, this is valid. Circuits produced by our algorithm are then minimized by simple template matching and identifying useless gates.